# Second and third harmonic nonlinear optical process in spray pyrolysed Mg:ZnO thin films

**Krithika Upadhya[1], Deekshitha U G[1], Albin Antony[1], Aninamol Ani[1], I.V Kityk[2], J.Jedryka[2], A. Wojciechowski[2], K.Ozga, Poornesh P[1*], Suresh D Kulkarni[3], N.Andrushchak[4]**

[1]Department of Physics, Manipal Institute of Technology, Manipal Academy of Higher Education, Manipal, Karnataka, 576104, INDIA

[2]Chair of Automatic, Electrotechnical and Optoelectronics, Faculty of Electrical Engineering, Czestochowa University of Technology, Armii Krajowej 17, PL-42-201 Czestochowa, Poland

[3]Department of Atomic and Molecular Physics, Manipal Academy of Higher Education, Manipal, Karnataka, 576104, INDIA

[4]Computer-Aided Design Department, Lviv Polytechnic National University, Lviv, Ukraine

Corresponding author: Poornesh P (poorneshp@gmail.com, poornesh.p@manipal.edu)

## Abstract

In the present study, Mg-doped ZnO (MZO) nanofilms (NF) were grown by spray pyrolysis technique. X-ray diffraction (XRD) patterns have shown an intense peak oriented along (002) crystalline plane confirming crystalline origin of grown MZO films. Atomic force microscopy (AFM) data revealed that the average roughness decreased significantly upon Mg doping from 20.3 nm to 4.72 nm. Ambient temperature photoluminescence spectra accounted for four emission bands in the visible spectral region due to the presence of various intrinsic defect centres in the films. The non-linear absorption coefficient ($\beta_{eff}$) were estimated by open aperture Z-scan technique and found to be in the order of $10^{-2}$ (cm/W) and indicates a positive absorption nonlinearity. Additionally, it was observed that incorporation of the metal Mg ions leads to a decay of the second harmonic generation (SHG) and to a non-regular increase of the third harmonic generation (THG).

**Keywords:** Mg: ZnO nanofilms, THG, Laser stimulated effects, Z-scan

## 1. Introduction

Recently, transparent conducting oxides (TCO's) exhibited enormous opportunities and attracted wide research interest because of their different topical characteristics and extensive range of applications. Among the widely studied TCOs, Zinc Oxide (ZnO) NF have always been exposed to a lot of studies due to their excellent combination of physical and chemical properties [1-4]. Relatively high optical transparency along with low electrical resistivity make ZnO NF a promising candidate for various optoelectronic devices such as solar cells [1], photovoltaic cells, flat panel display [2], gas sensors [3], light emitting diodes etc. Novel approaches and growth techniques are being developed in order to enhance and control the physical and chemical features of ZnO films to be suitable for practical applications[5-7]. Among the widely adopted methods, one of the promising ways is to doping with suitable elements effectively. The selection of dopant element having a comparable radius with a host atom is most appropriate as it can reduce lattice distortion. Hence, In the present study we have chosen Magnesium (Mg) as a dopant element whose ionic radius (~0.57 Å) was comparable with $Zn^{2+}$ (0.60 Å) [4]. Consequently, Mg ion can alter both the physical and chemical properties of ZnO without inducing lattice distortions.

The unique combination of physical and chemical properties makes ZnO thin films compatible to various synthesis and deposition techniques [5-9]. The selection of synthesis methods depends on the required morphology and applications of the films [7-10]. Thin film deposition methods like Pulsed laser deposition, DC magnetron sputtering, sol-gel method, electrodeposition, chemical vapour deposition, spray pyrolysis etc. are employed for the growth of ZnO NF. Here, we implemented spray pyrolysis technique which is a low cost, simple and flexible technique compared to others. The ability of large area deposition in a short time with good quality thin films makes them a widely accepted method for depositing thin film in industrial applications. Magnesium doped ZnO NF were deposited with different proportion of Mg doping and its variations induced on structural, optical and morphological properties were elaborated by atomic force microscopy (AFM) operated in tapping mode configuration, X-Ray diffraction (XRD), UV visible [4] and photoluminescence (PL)spectroscopy [11]. Furthermore, ZnO NF shows an excellent nonlinear optical response both in second and third order level which have multiple applications in various optoelectronic and nonlinear optical device domain [12]. Interestingly, the studies pertaining to second and third harmonic generation of ZnO NF and MZO thin nanofilms were not explored till date and requires further investigation.

In this context, an attempt to study harmonic generation in MZO thin nanofilms at various Mg doping concentration were made. The nonlinear optical measurements were made in both pulsed and continuous wave regime and elaborated [13]. The credibility of the grown NF in frequency conversion and optical limiting application were also discussed in brief.

## 2. Experimental Procedure

*2.1 Fabrication of MZO thin films*

ZnO NF of both pure and MZO were grown using spray pyrolysis technique. The starting solution of Zinc chloride [$ZnCl_2$] was prepared by mixing 50ml of double distilled water and 50ml of Ethanol. Doping is achieved by adding of Magnesium chloride hexahydrate salt [$MgCl_2.(H_2O)_6$] dissolved in 50 ml of distilled water at ambient temperature. The solution with a molarity of 0.05M is prepared. The concentration of Mg dopant was varied from 5 wt.% to 15wt.%. The deposition temperature was maintained at 400$^0$C with a fixed flow rate of 2 ml/min. The thickness of grown films was investigated using Bruker Stylus profilometer and found to be 300 nm.

*2.2 Characterizations of MZO thin films*

Structural properties of MZO thin films were determined using high-resolution X-ray diffraction using CuKα (0.14506nm) radiation. Variation in optical transmittance and band gap energy were studied with a spectral resolution of 1 nm by UV-VIS spectrophotometer [14]. Surface morphology of the grown NF was analysed by AFM in a scan area of 5μm. The root mean square roughness parameter was estimated from Bruker nanoscope analysis software. Photoluminescence spectroscopy technique was adopted with the intention to understand the electronic structure of the sample and to examine the radiative transitions in MZO thin films. Non-linear absorption properties of the fabricated films were scanned through Z-scan technique under continuous wave laser illumination. The photo induced nonlinear optical studies were performed using fundamental 10 ns Nd: YAG laser and the nonlinear optical signals were detected as per the procedure described in the reference [15].

## 3. Results and discussion

### 3.1 Structural properties

The variation in structural properties was determined by analysing the XRD peaks obtained for NF deposited at different concentration of Mg. The XRD pattern of pure and MZO NF are depicted in Fig. 1. Undoped ZnO NF exhibit a polycrystalline growth phase indicated by multiple peaks oriented along (100), (002), (101), (102), (110), (103), (112) and (201). For the MZO films, a sharp (002) peak was dominant [16] and peak intensity is observed to be increased with doping. The peak intensity was maximum for Mg dopant concentration equal to about 15 wt% confirming the improvement in the crystallinity upon doping. The presence of single phase in the XRD pattern confirms the replacement of $Zn^{2+}$ by $Mg^{2+}$ which has similar ionic radii (0.72A) and does not change its structure [17]. The crystallite size 'D' (in nm) of MZO thin film for (002) peak was estimated using Debye-Scherer equation and shown in table 1. [18]. It is observed that crystallite size shows an enhancement from 13.30 nm to 43.81 nm upon Mg incorporation hence indicates the enhancement in crystallinity of the nanostructures. The structural parameters such as strain and dislocation density of MZO NF were also evaluated [18] and presented in Table 1. The strain and dislocation found to be decreasing for the doped ZnO films which indicate the superior quality of films grown using the chemical route.

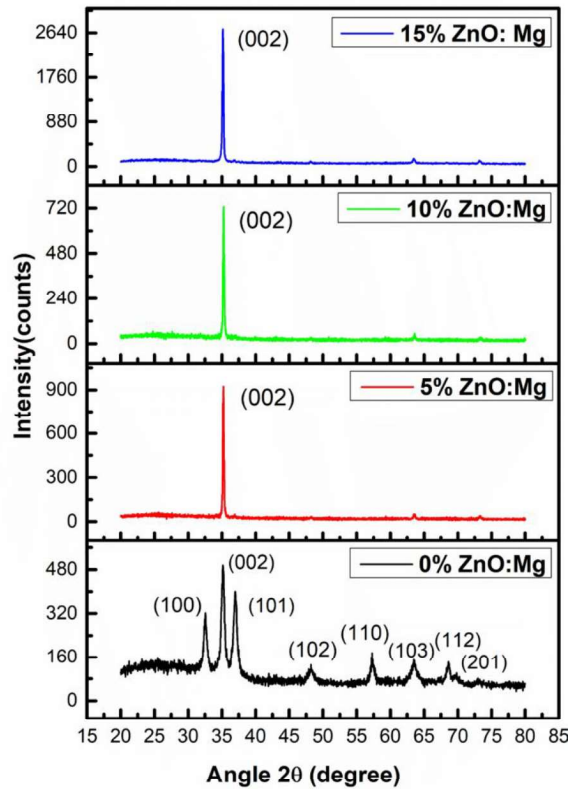

**Fig. 1.** XRD Pattern for MZO thin films

Table 1. Structural parameters of MZO thin films

| Dopant Mg (wt %) | FWHM (degree) | 2θ (degree) | Crystallite size D (nm) | Dislocation density δ ×10$^{14}$/m | Strain (×10$^{-3}$) ε |
|---|---|---|---|---|---|
| 0 | 0.62 | 35.1 | 13.30 | 56.45 | 8.63 |
| 5 | 0.19 | 35.27 | 42.66 | 5.47 | 2.67 |
| 10 | 0.18 | 35.27 | 43.81 | 5.17 | 2.60 |
| 15 | 0.20 | 35.14 | 40.87 | 5.98 | 2.80 |

### 3.2 Morphological Properties

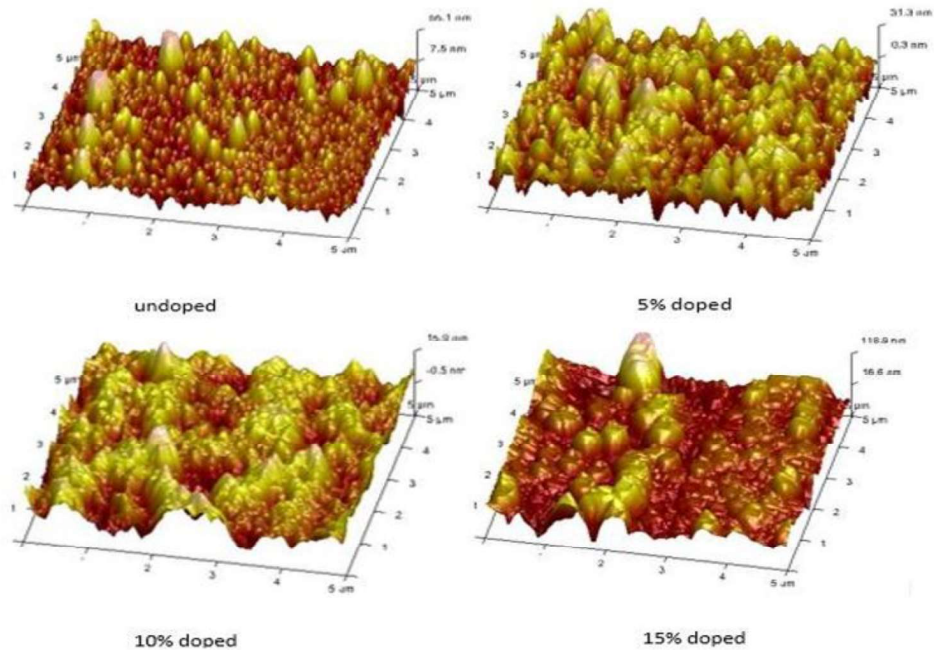

**Fig. 2.** AFM micrographs of MZO thin films

Fig. 2 shows the 3D AFM micrographs of pure and MZO thin films. Incorporation of Mg leads to a significant variation in the surface morphology of the films grown. The surface roughness of pure ZnO and MZO films was evaluated and tabulated in Table 2. The distribution of small spherical shaped grains observed in the undoped ZnO NF is shown in Fig 2(a). Substantial reduction in surface roughness from 20.3 nm to 4.72 nm was observed with an increase in the doping of Mg from 0% to 10 % indicates that scattering is less and it is expected to exhibit good nonlinear optical property

**Table 2.** Variation in the roughness of MZO thin films

| Dopant Mg (wt.%) | Roughness (nm) |
|---|---|
| 0 | 20.3 |
| 5 | 8.81 |
| 10 | 4.72 |
| 15 | 12.7 |

### 3.3 Linear optical properties

Transmission spectra are shown in the Fig.3 (a) and show higher level of transmittance in the visible spectrum due to Mg addition compared to that of undoped ZnO. The wave-like patterns observed were due to the interference effects at various interfaces between the substrate, air and sample. A sharp cut off in transmittance was observed for ultraviolet spectral range indicating that the fabricated MZO films can be employed for UV detection [4]. The spectral shifting of absorption edge towards the lower wavelength of the spectrum are due to Burstein Moss effect. The shift indicates the variation in energy band gap due to the assimilation of Mg in ZnO lattice [19, 20]. The band gap of pure and doped ZnO films was calculated using Tauc's plot analysis and depicted in Table 3. The values were observed to be increased from 3.25eV-3.32eV due to Burstein Moss effect.

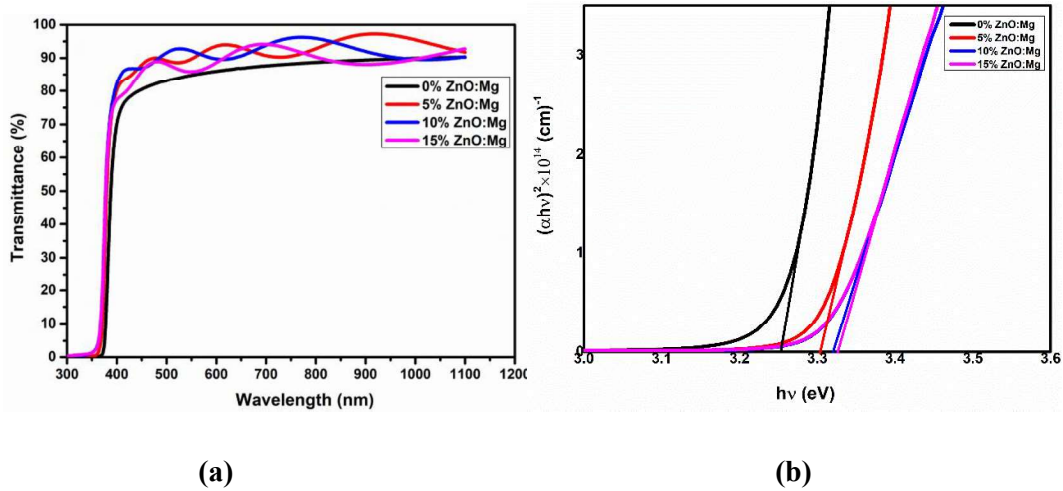

**Fig. 3.** (a) Transmittance vs Wavelength (b) (αhν)² vs hν plot of MZO thin films

**Table 3.** Variation in band gap of MZO films at different Mg doping conc.

| DopantMg (wt.%) | Band gap (eV) |
|---|---|
| 0 | 3.25 |
| 5 | 3.30 |
| 10 | 3.31 |
| 15 | 3.32 |

### 3.4 Photoluminescence properties

Incorporating of Mg into ZnO lattice results in occurrence of number of defects that can change linear and non-linear optical properties [21]. Intrinsic defects like Zn interstitials, oxygen or zinc vacancy defects could be formed and act as recombination centres or trap states within the forbidden gap [22]. In order to study such defect states, ambient temperature photoluminescence spectra of the NF were studied by exciting at a wavelength of 345nm [23] and are shown in the Fig 4. Photoluminescence spectral intensity was found to be enhanced with Mg concentration which infers that radiative transitions in the films shows an enhancement upon Mg incorporation. The enhancement in the observed radiative transitions confirm presence of various radiative defect states. The obatined PL curves were deconvoluted using Gaussian fit (see Fig. 5) to detect the origin of defect states. The spectra were broad and consisted of four emission bands along violet, blue, green and red colour centre in the visible spectral range [21]. For green emission at ~2.23eV, were due to singly ionised oxygen vacancy [24]. A prominent luminescence peak at ~2.6 eV corresponded to a blue colour centre were caused by the zinc interstitials defect site. At the same time the violet at ~3 eV and red emissions at ~1.9 eV have been evidently arising due to the presence of oxygen antisite and oxygen interstitials defect sites respectively.

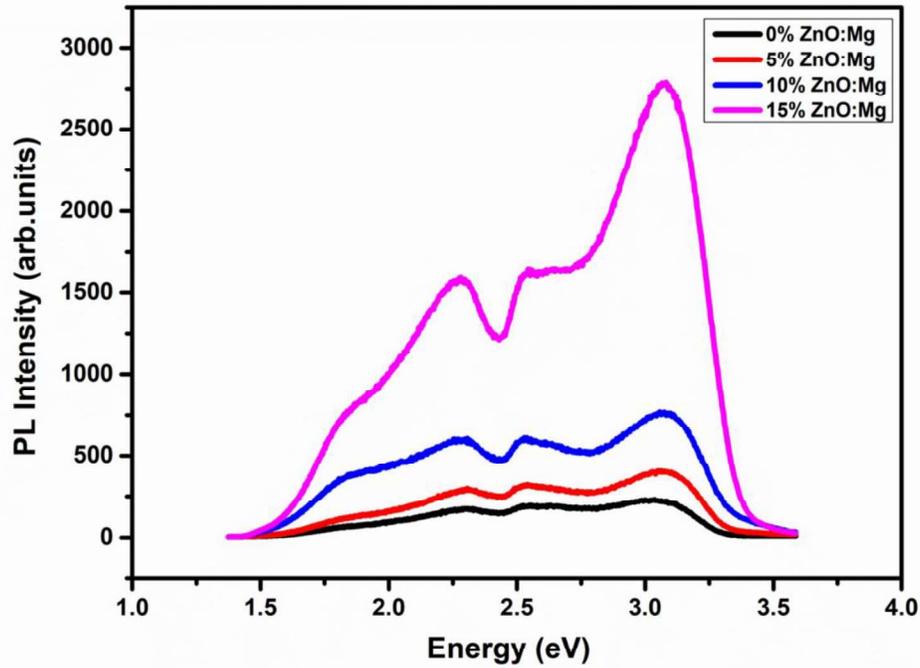

Fig. 4. PL spectra of MZO thin films.

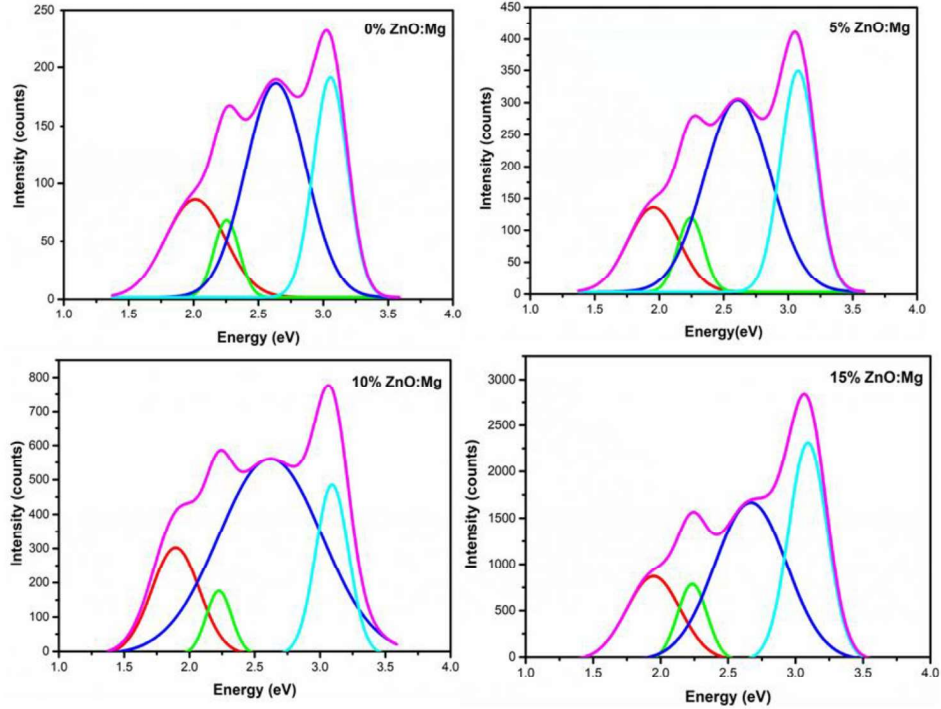

Fig. 5. Deconvoluted PL spectra MZO thin films

**3.5 Non-linear optical properties**

3.5.1 Open aperture Z-scan studies:

The open aperture Z-scan technique proposed by Sheikh Bahae et al.[25] was carried out by using continuous wave He-Ne laser of wavelength 632nm and input power of 22mW. The trace of open aperture Z-scan in Fig. 6 shows reverse saturable absorption (RSA) behaviour. This accounts for the positive absorption nonlinearity. In semiconductors, RSA behaviour is attributed to the presence of various mechanisms such as two-photon absorption (TPA), induced scattering, free carrier absorption etc. In the present study, TPA is allowed only if the incident energy ℏω is higher that $E_g/2$ but less than band gap energy $E_g$ [6]. In MZO films this criterion was fulfilled by the incident energy used for excitation [23]. In the present case of MZO NF the exhibited nonlinear absorption is attributed to FCA induced two-photon absorption since the nonlinearity obtained belong to resonant nonlinear mechanism [23].

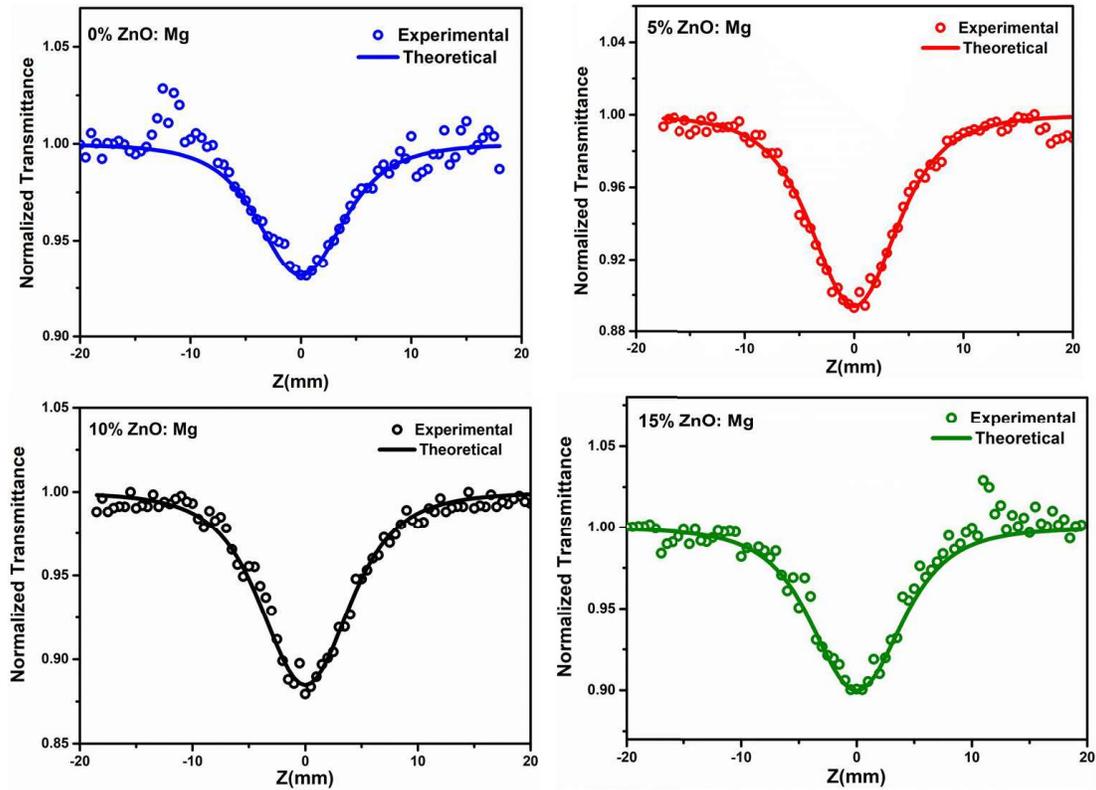

Fig. 6. Open aperture Z-scan traces of MZO thin films

**Table 4.** Non-linear absorption coefficient obtained for MZO thin films

| Samples | $\beta_{eff}$ ($\times 10^{-2}$ cm/W) |
|---|---|
| 0% ZnO: Mg | 6.24 |
| 5% ZnO: Mg | 9.72 |
| 10% ZnO: Mg | 11.01 |
| 15% ZnO: Mg | 9.11 |

3.5.2 Laser stimulated second and third harmonic generation studies:

The measurements of the laser stimulated SHG and THG have been performed following the similar method described in the ref. [26]. The principal experimental setup is shown in Fig. 7. The NF have been laser treated by two coherent beams with coherent frequencies and beam spots diameter equal to about 15 mm for 2-3 min. Additionally the scattered background waves were monitored by spherical quantum sphere. The treatment was performed several minutes until a saturation of the laser induced absorption was achieved. Principally it was limited by the appearance of photodestruction of the samples. The output SHG has been measured by green interference filters at 532 nm, 520 nm and 540 nm and 305 nm, 315 nm, 325 filters were used for THG signal detection. These filters were necessary to eliminate parasitic fluorescence background. Only the half order spectral jump of the THG and SHG signals were necessary for confirmation of nonlinear optical effects.

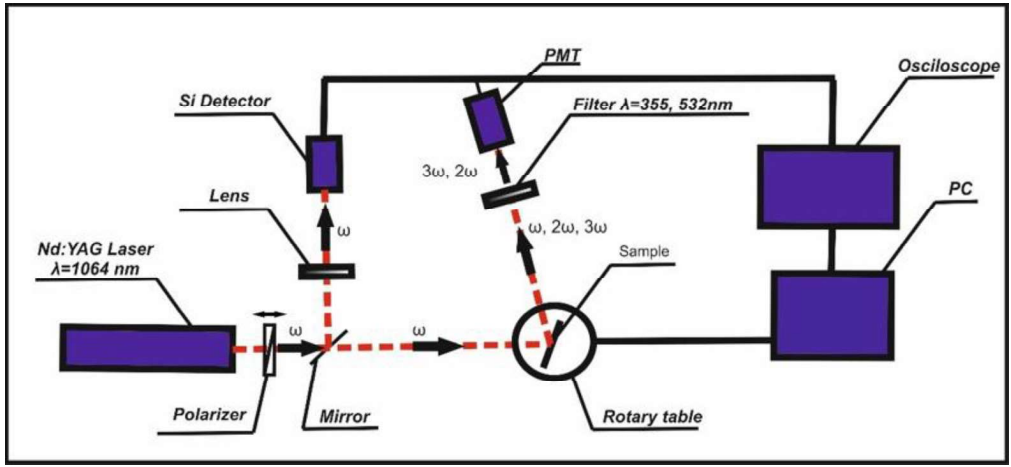

Fig. 7. General set-up for the SHG and THG measurements,

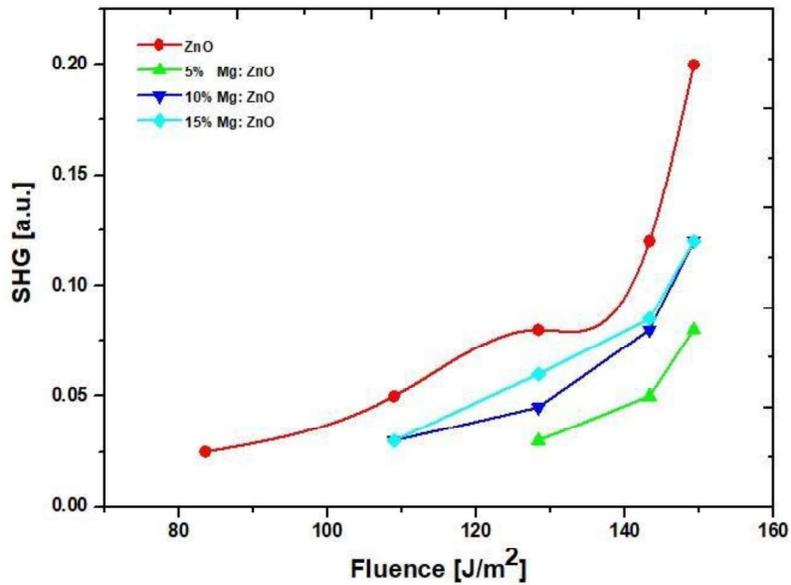

Fig.8. Laser stimulated SHG efficiency of MZO thin films versus the fundamental laser energy.

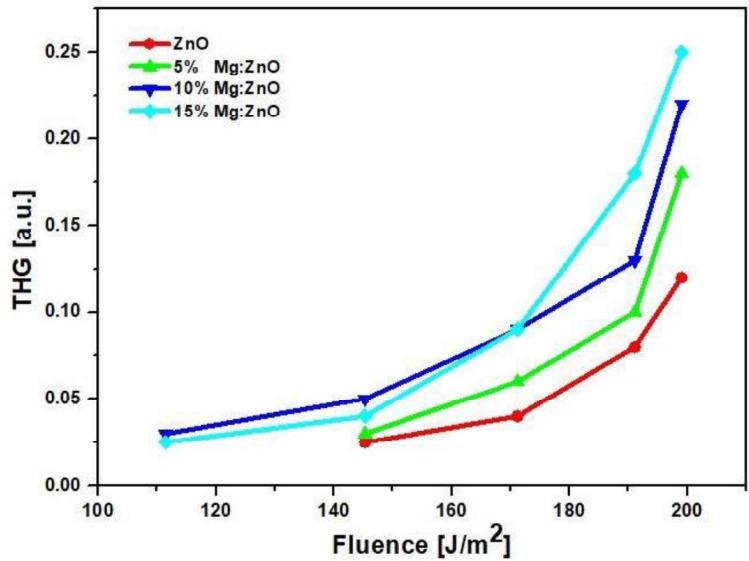

Fig.9. Laser stimulated THG spectra of MZO thin films

The dependence of SHG and THG signal intensity versus the fundamental beam is different for the second and third harmonic generation as shown in Fig. 8 and Fig. 9. The typical dependences of THG (Fig. 9) versus the fundamental laser beams have clearly shown that the incorporation of the metal Mg ions leads to decay of the SHG with non-monotonous dependence and to a non-regular increase of the THG. At the same time, 5 % of Mg show the minimal signals both for the SHG as well as the THG. Such principal different dependence may be explained following the origin of these two nonlinear optical effects. In the case of the SHG principal role begins to play by space charge density eccentricity [27] and for the THG the principal role is played by the changes of the dipole moments of the excited states [28].

## Conclusions

The nonlinear optical responses of second and third order nonlinear optical efficiencies with respect to the Mg content were reported for ZnO Mg crystalline nanofilms. The laser stimulated SHG shows a decrease with the addition of the Mg and the THG shows quite opposite behaviour. All the dependencies demonstrate non-monotonous Mg content dependencies. It may be partially determined by the charge density eccentricity as well as by morphological parameters. With the incorporation of Mg in ZnO nanostructures, a decrease in the roughness was observed which indicates that smoother films were formed with scattering free surface. Enhanced transmittance within the visible region and cut off at the ultraviolet region leads to the possibility of MZO films to be employed for UV detection. A preferential growth orientation along (002) plane observed on Mg doping along with an increase in crystallite size accounts to the growth of high-quality crystalline films. Photoluminescence studies validate the oxygen defect states formation in the films. Non-linear absorption coefficient values indicate the MZO films showsa positive absorption nonlinearity.

## Acknowledgements

Concerning N.A. and I.V.K., A.W.J.,presented results are part of a project that has received funding from the European Union's Horizon 2020 research and innovation programme under the Marie Skłodowska-Curie grant agreement No 778156.